# Growth, structure, micro-structure and magneto transport of an easy route synthesized bulk polycrystalline TiSe$_2$

Abhilasha Saini[1,2], Kapil Kumar[1,2], M.M. Sharma[1,2], R.P. Aloysius[1,2] and V.P.S. Awana[1,2]

[1]*CSIR-National Physical Laboratory, Dr. K. S. Krishnan Marg, New Delhi-110012, India*
[2]*Academy of Scientific and Innovative Research (AcSIR), Ghaziabad 201002, India*

**Abstract:**

This article reports an easy route synthesis of bulk polycrystalline TiSe$_2$. Phase purity and microstructure are determined through powder X-ray diffraction (PXRD) and field emission scanning electron microscopy (FESEM) respectively. Vibrational modes of TiSe$_2$ as being analyzed by Raman spectroscopy, show the occurrence of both A$_{1g}$ and E$_g$ modes. Charge density wave (CDW) is observed in transport measurements of TiSe$_2$ with a hysteresis in cooling and warming measurements at around 180K. Further, studied TiSe$_2$ showed negative magneto resistance (MR) below the CDW and a small positive MR above the CDW.

*Corresponding Author

Dr. V. P. S. Awana: E-mail: awana@nplindia.org
Ph. +91-11-45609357, Fax-+91-11-45609310
Homepage: awanavps.webs.com

**Introduction:**

Transition metal dichalcogenides (TMDCs) are a class of layered materials, having an atomic layer of transition metal being sandwiched between two atomic layers of chalcogen atoms. The adjacent layers of TMDCs are stacked via vander waals forces [1-3], and thus qualifying them to be the members of van der Waals compounds family [4-6]. TMDCs are metallic, semi-metallic or semiconducting, depending on the choice of the insertion of transition metal and chalcogen [7]. This class includes, semiconductors such as MoS$_2$, WS$_2$ [8], and semimetals e.g. WTe$_2$, TiSe$_2$ [9,10]. Also, some TMDs like NbSe$_2$ are found to show superconductivity as well [11]. TiSe$_2$ is one of the most studied materials in TMDCs family. F.J. Di Salvo et al. first reported a commensurate 2×2×2 charge modulation or CDW in TiSe$_2$ at 200K accompanied by a periodic lattice distortion [12]. The theoretically calculated bulk electronic band structure showed that TiSe$_2$ must be a semimetal, with non-trivial topological properties [13]. Due to the presence of the both semi metallic nature and topological properties, TiSe$_2$ is regarded as a Topological semimetal (TSM) [13]. TiSe$_2$ is found to show direct correlation between observed CDW and bulk electronic band structure as being evidenced by angle resolved photo-electron spectroscopy [14].

There have been several studies on CDW characteristics of TiSe$_2$ [15-18]. CDW is known to get suppressed with doping or intercalation of foreign atoms [15-18]. Apart from doping of foreign atoms, CDW can be altered by different growth mechanism or by surface oxidation [19,20]. TiSe$_2$ is also found to show Kondo effect on doping of 3d transition elements, which acts as a spin flip centres in the TiSe$_2$ unit cell [15]. More interestingly, TiSe$_2$ is found to show superconductivity with doping of Cu atoms and Pd atoms [16,21], this result was interesting in itself as the observed superconductivity was accompanied by suppressing of CDW with increased doping. Although, CDW is the most prominent feature of TiSe$_2$, yet its origin is different from most of the materials showing the same property. In most of the material CDW emerges from Fermi surface nesting, while the same reason is not valid in case of TiSe$_2$ [16]. All in all, TiSe$_2$ is an interesting material with some exceptional properties. This motivates us to write this short letter on an easy route synthesis and brief physical properties of TiSe$_2$.

In this short letter, we report synthesis of polycrystalline TiSe$_2$. The synthesized sample is examined in context of phase purity and elemental composition through XRD and energy dispersive X-ray analysis (EDAX) techniques respectively. Raman modes are also observed for polycrystalline TiSe$_2$ through Raman spectroscopy. The signature feature of TiSe$_2$ i.e., CDW is observed at around 180K, with a hysteresis in warming and cooling data. Magneto-transport measurements shows positive MR above CDW temperature and the same is negative below the CDW temperature.

**Experimental:**

Polycrystalline samples of TiSe$_2$ were prepared using an easy self-flux method. The stoichiometric amounts (1:2) of Ti (99.99 %) and Se (99.99 %) powders were taken and ground thoroughly in MBRAUN glove box in argon atmosphere. The obtained mixed powder was pressed into a rectangular pellet using the hydraulic press under an approximate pressure of 2 gm/cm³ and were sealed in a quartz tube under a vacuum of $5\times10^{-5}$ Torr. The vacuum quartz tube was placed into PID controlled muffle furnace, heated to 650°C at a rate of 120°C/h, at this point Ti atoms reacts with Se atoms. After this, the obtained pre-reacted powder was grounded again and heated to a higher temperature of 950°C at a ramp rate of 120°C/h. Sample is heated at a higher ramp rate to minimize the loss of Se. The sample was kept at 950°C for 48 hours and subsequently cooled down to 650°C with rate of 2°C/h. During this period $TiSe_2$ phase start to develop and this phase is stabilized by keeping the sample at 650°C for 24 hours and was finally cooled to room temperature. Fig. 1 (a) & (b) show the schematic of both of the stages of synthesis process of polycrystalline $TiSe_2$.

The structural analysis for checking the phase purity was done through room temperature XRD (X-ray diffraction) using Rigaku Miniflex II X-ray diffractometer equipped with Cu-Kα radiation of wavelength 1.5418Å. Full prof software is used for Rietveld refinement of PXRD pattern and VESTA software is used to extract unit cell based on refined parameters from Rietveld analysis. Surface morphology and EDAX spectra were taken using Zeiss EVO-50 FESEM. Raman spectra of bulk $TiSe_2$ polycrystalline is taken at room temperature using the Jobin Yuvon Horiba T64000 Raman Spectrometer for a wavenumber range of 55-400 $cm^{-1}$. Magneto-transport measurements were carried out using Quantum Design Physical Property Measurement System (QD-PPMS), a standard four probe method is used to measure the transport properties of synthesized $TiSe_2$ polycrystalline sample.

**Result & Discussion:**

Fig. 2(a) shows Rietveld refined PXRD pattern of synthesized $TiSe_2$ polycrystalline sample. Rietveld refined PXRD pattern confirms that the synthesized $TiSe_2$ sample is crystallized in trigonal crystal structure with P -3 m 1 (164) space group. The synthesized sample seems to be phase pure through PXRD pattern as all the peaks are found to be indexed with their corresponding planes in $TiSe_2$ crystal structure. No XRD peak corresponding to any impurity peak is detected. The quality of fit is determined through calculating goodness of fit parameter i.e. $\chi^2$ parameter, which is found to be 5.63. The Rietveld refined lattice parameters along with the atomic positions are tabulated in Table-1. The unit cell of $TiSe_2$ is drawn by using VESTA software, which employs crystallographic information file generated through Rietveld refinement of PXRD pattern. The same is shown in Fig. 2(b), it is clear that the studied $TiSe_2$ polycrystalline sample exhibits a layered structure composed of van der Waals layers. Each van der Waals layer contains tri-layers arranged in Se-Ti-Se order, in which the Ti atom is surrounded by six Se atoms in the octahedral configuration. The presence of van der Walls gap in between the trilayer makes the compound easily cleavable and provides opportunity to the host and the dopant material for various intercalations and the possible charge transfer etc.

FESEM image of synthesized $TiSe_2$ sample is shown in inset (a) of Fig. 3. The elemental composition is determined through EDAX measurements and the same is shown in inset (b) of Fig. 3. Both the constituent elements viz. Ti and Se are found to be in a near stoichiometric ratio, signifying the purity of the sample. Also, the EDAX spectra shown in Fig. 3 shows the peaks only for Ti and Se atoms, which shows that the synthesized sample is free from any foreign contamination. Fig. 4 presents the room temperature Raman spectra of synthesized bulk $TiSe_2$ polycrystalline sample. Raman vibration modes are observed at $150 cm^{-1}$ and $255 cm^{-1}$ respectively. These vibrational modes are identified as $A_{1g}$ and $E_g$ modes of $TiSe_2$ and are in accordance with the previous report on 1T phase of $TiSe_2$ [22]. In 1T phase of TMDs, transition element is octahedrally bonded with the chalcogen atoms. The schematic of vibrational modes observed in $TiSe_2$ are shown in inset of Fig. 4. $A_{1g}$ modes of $TiSe_2$ comprises in-plane vibrations of Se atoms, while the $E_g$ modes show out of plane vibrations of Se atoms around Ti atoms. This is clear from inset of Fig. 4.

Fig. 5(a) shows ρ-T measurements result of $TiSe_2$ polycrystalline samples. An interesting anomaly, in terms of insulator to metal transition is observed in a temperature range of 150K-200K. Charge carriers gets ordered and thus the observance of CDW at around 180K. This result is in accordance with the earlier reported results on the same compound [15-20]. Further, clear hysteresis is observed in CDW in both warming and cooling data, which is the characteristic of a first order phase transition. The reason behind this hysteresis is vested in dual electrical conductivity of $TiSe_2$. $TiSe_2$ shows two electronic phases, namely at higher temperatures i.e., above CDW the insulating phase starts to dominate over the metallic phase across CDW, as revealed in Fig. 5(a). At CDW this insulting phase is maximized but the metallic phase does not disappear. The coexistence of two electronic phases results in the hysteresis in warming and cooling cycle.

Fig. 5(b) shows MR% vs applied field (up to ±10T) plot of synthesized $TiSe_2$ sample at 2K, 100K, 200K and 290K. MR% is calculated by following formula

$$MR\% = \frac{[\rho(H)-\rho(0)]}{\rho(0)} \times 100$$

MR% is calculated at two temperature points below the CDW transition viz. 2K and 100K and same way above CDW transition i.e., at 200K and 290K. MR% behaviour is different in both the regions. TiSe$_2$ is found to show positive MR at temperatures above the CDW transition and a negative MR at temperatures below the same. The negative MR is found to be around 3.5% at 2K and 1% at 100K. This negative MR is related to the correlation between CDW and magnetic field. CDW tends to suppress with applied magnetic field and eventually demolished when Pauli spin energy surpasses the CDW condensation energy. This argument has been suggested to be the reason for observed negative MR below CDW transition [23,24]. Another possible reason for negative MR below CDW could be the possible Kondo effect, however that is not the case here, as in Kondo effect MR shows parabolic dependence on applied magnetic field and not linear [25]. Above CDW, MR is positive but very small around 0.4%. The possible reason for the observed behaviour of MR can be related to simultaneous existence of two electronic phases in TiSe$_2$. The insulating electronic phase tends to increase gradually as the temperature is reduced down to CDW transition, and this insulating electronic phase of TiSe$_2$ results in small but a positive MR above CDW transition. But at the temperatures below the CDW transition, metallic phase tends to increase with lowering the temperature and a negative MR is observed below CDW transition. In insulating region MR is very low (<1%) and nearly constant at both the temperatures viz. 200K and 290K, which is obvious as the insulating phase of a material shows much lesser effect of magnetic field on resistivity. While, in metallic phase, the observed negative MR shows stronger temperature dependency, as it is increased to 3.5% at 2K from 1% that was at 100K. This stronger temperature dependency of MR below CDW transition is due to metallic phase of TiSe$_2$.

**Conclusion:**

In this work, we synthesized TiSe$_2$ polycrystalline sample, which is well characterized through PXRD, FESEM and Raman spectroscopy. Raman spectroscopy confirms 1T phase of TiSe$_2$. This 1T TiSe$_2$ is found to show ordering of carriers at around 180K in terms of CDW, which further show hysteresis in warming and cooling cycles. Magnetoresistance measurements show interesting behaviour as the synthesized sample is found to show positive MR above CDW transition and negative MR below CDW transition. Our motive through this short letter is to inspire the condensed matter scientists to study more about the MR properties at above and below CDW transition.

**Acknowledgement:**

The authors would like to thank the Director of National Physical Laboratory (NPL), India, for providing the facilities and his keen interest in research. Abhilasha Saini, M. M. Sharma and Kapil Kumar would like to thank AcSIR-NPL for Ph.D. registration.

**Table 1**

**Cell parameters of TiSe$_2$**

| a(Å) | b(Å) | c(Å) | α | β | γ |
|---|---|---|---|---|---|
| 3.549(2) | 3.549(2) | 6.011(4) | 90° | 90° | 120° |
| Atom | Wyckoff | x | y | Z | Occupancy |
| Ti | 1a | 0.000 | 0.000 | 0.000 | 1 |
| Se(I) | 4d | 0.3333 | 0.6666 | 0.231(1) | 1 |
| Se(II) | 12j | 0.6666 | 0.3333 | 0.768(8) | 1 |

**References:**

1. Alexander A. Balandin, Sergei V. Zaitsev-Zotov and George Gruner, Appl. Phys. Lett. **119**, 170401 (2021).
2. Davied B. Lioi, David J. Gosztola, Garry P. Wiederrecht and Goran Karapetrov, Appl. Phys. Lett. **110**, 081901 (2017).
3. Sajedeh Manzeli, Dmitry Ovchinnikov, Diego Pasquier, Oleg V. Yazyev & Andras Kis, Nat Rev Mater **2**, 17033 (2017).
4. A.N. Titov, Yu. M. Yarmoshenko, P. Bazylewski, M.V. Yablonskikh et al. Chemical Physics Letters **497,** 187 (2010).
5. M.B. Dines. Science **188**, 1210 (1975).
6. C. H. Lee et al., Nat. Nanotechnol. **9**, 676 (2014).
7. Q. H. Wang, K. Kalantar-Zadeh, A. Kis, J. N. Coleman and M. S. Strano, Nat. Nanotechnol. **7**, 699 (2012).
8. Maryam Nayeri, Mahdi Moradinasab and Morteza Fathipour, Semicond. Sci. Technol. **33**, 025002 (2018).
9. Yuqiang Li, Jingxia Liu, Peiguang Zhang, Jianxin Zhang, Ningru Xiao, Liyuan Yu & Pingjuan Niu, J Mater Sci **55**, 14873 (2020).
10. Julia C. E. Rasch, Torsten Stemmler, Beate Müller, Lenart Dudy, and Recardo Manzke, Phys. Rev. Lett. **101**, 237602 (2008).
11. H. N. S. Lee, H. McKinzie, D. S. Tannhauser, and A. Wold, J of App. Phys. **40**, 602 (1969).
12. F. Di Salvo, Jr., D. E. Moncton, and J. V. Waszczak, Phys. Rev. B 14, 4321 (1976).
13. Shin-Ming Huang, Su-Yang Xu, Bahadur Singh, Ming-Chien Hsu, Chuang-Han Hsu, Chenliang Su, Arun Bansil and Hsin Lin, New J. Phys. **23**, 083037 (2021).
14. M.-L. Mottas, T. Jaouen, B. Hildebrand, M. Rumo, F. Vanini, E. Razzoli, E. Giannini, C. Barreteau, D. R. Bowler, C. Monney, H. Beck, and P. Aebi. Phys. Rev. B **99**, 155103 (2019).

15. M. Sasaki, A. Ohnishi, T. Kikuchi, M. Kitaura, Ki-Seok Kim, and Heon-Jung Kim, Phys. Rev. **B** 82, 224416 (2010).
16. E. Morosan, H. W. Zandbergen, B. S. Dennis, J. W. G. Bos, Y. Onose, T. Klimczuk, A. P. Ramirez, N. P. Ong, and R. J. Cava, Nature phys. **2**, 544 (2006).
17. D. Qian, D. Hsieh, L. Wray, E. Morosan, N. L. Wang, Y. Xia, R. J. Cava, and M. Z. Hasan, Phys. Rev. Lett. **98**, 117007 (2007).
18. P. Behera, Sumit Bera, M. M. Patidar, and V. Ganesan, AIP Conference Proceedings **2100**, 020113 (2019).
19. Lifei Sun,Chuanhui Chen,Dr. Qinghua Zhang,Christian Sohrt,Tianqi Zhao,Dr. Guanchen Xu,Jinghui Wang,Prof. Dong Wang,Prof. Kai Rossnagel,Prof. Lin Gu,Prof. Chenggang Tao,Prof. Liying Jiao, Angewandte Chemie **56**, 8981 (20190.
20. Matthias M. May, Christine Brabetz, Christoph Janowitz, Recardo Manzke, Journal of Electron Spectroscopy and Related Phenomena **184**, 180 (2011).
21. E. Morosan, K. E. Wagner, Liang L. Zhao, Y. Hor, A. J. Williams, J. Tao, Y. Zhu, and R. J. Cava, Phys. Rev. B **81**, 094524 (2010).
22. Ranu Bhatt, Miral Patel, Shovit Bhattacharya, Ranita Basu, Sajid Ahmad, Pramod Bhatt, A K Chauhan, M Navneethan, Y Hayakawa, Ajay Singh, D K Aswal and S K Gupta, J. Phys. Cond. Mat. **26**, 445002 (2014).
23. H. S. J. van der Zant, E. Slot, S. V. Zaitsev-Zotov, and S. N. Artemenko, Phys. Rev. Lett. **87**, 126401 (2001).
24. Hongjun Xu, Ming-Chien Hsu, Huei-Ru Fuh, Jiafeng Feng, Xiufeng Han, Yanfeng Zhao, Duan Zhang, Xinming Wang, Fang Liu,Huajun Liu, Jiung Cho, Miri Choi, Byong Sun Chun, Cormac Ó Coileáin, Zhi Wang, Mansoor B. A. Jalil, Han-Chun Wu and Ching-Ray Chang, J. Mater. Chem. C **6**, 3058 (2018).
25. Y. Katayama, & S. Tanaka, Phys. Rev. **153**, 873 (1967).

**Figure Captions:**

**Fig. 1(a)** Schematic of 1st step of heat treatment followed to synthesize TiSe$_2$ polycrystalline sample. **(b)** Schematic of 2nd step of followed heat treatment.

**Fig. 2(a):** Rietveld refined PXRD of polycrystalline TiSe$_2$ sample **(b)** Unit Cell Structure through VESTA software.

**Fig. 3:** EDAX spectra of synthesized TiSe$_2$ polycrystalline sample in which inset (a) is showing the FESEM image and the inset (b) is showing the atomic composition of constituent elements.

**Fig. 4:** Room Temperature Raman Spectra of TiSe$_2$ polycrystalline in which inset is showing the schematic of TiSe$_2$ Raman Modes (A$_{1g}$ & E$_g$) modes.

**Fig. 5(a):** ρ-T measurements of TiSe$_2$ polycrystalline sample in warming and cooling cycles **(b)** MR% vs H (±10T) plots of TiSe$_2$ polycrystalline sample at 2K, 100K, 200K and 290K.

**Fig.1(a)&(b)**

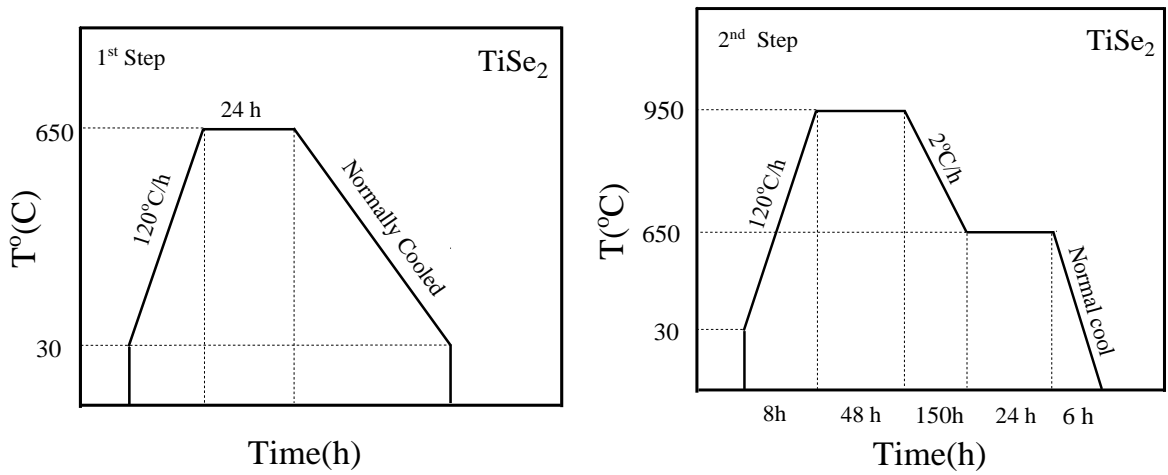

**Fig. 2**

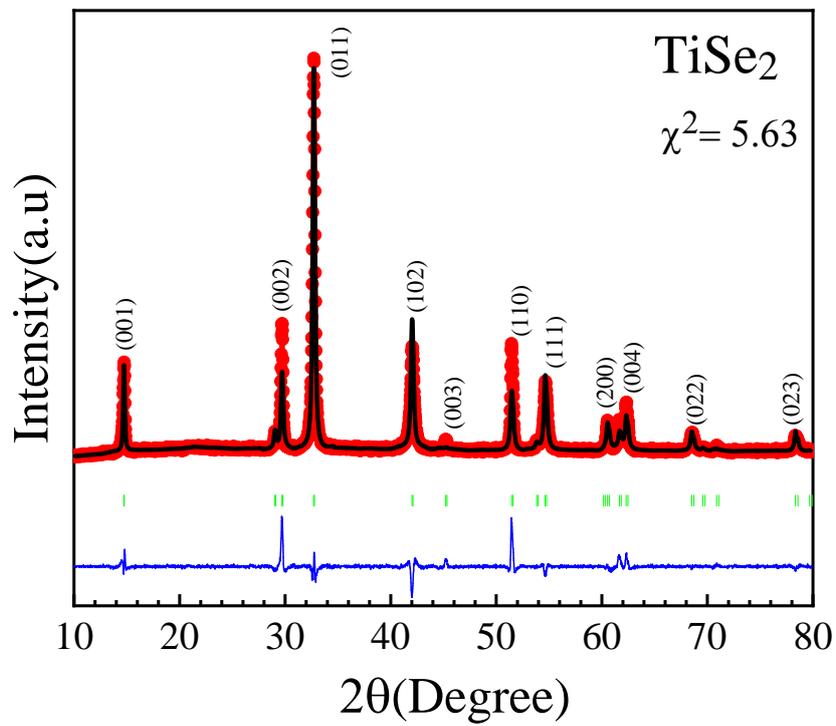

**Fig. 2(b)**

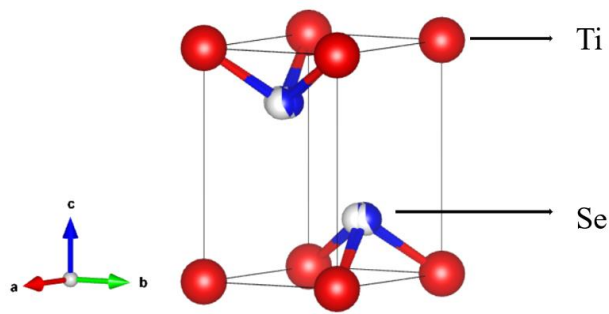

**Fig. 3**

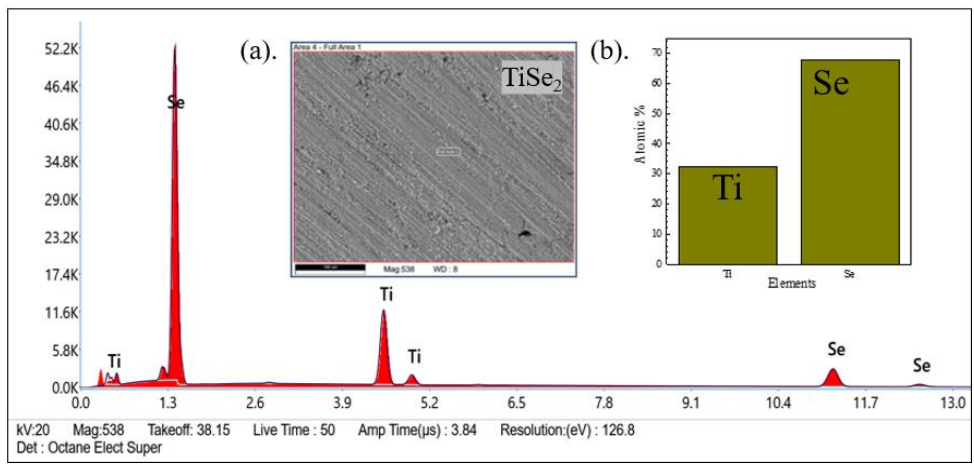

**Fig. 4**

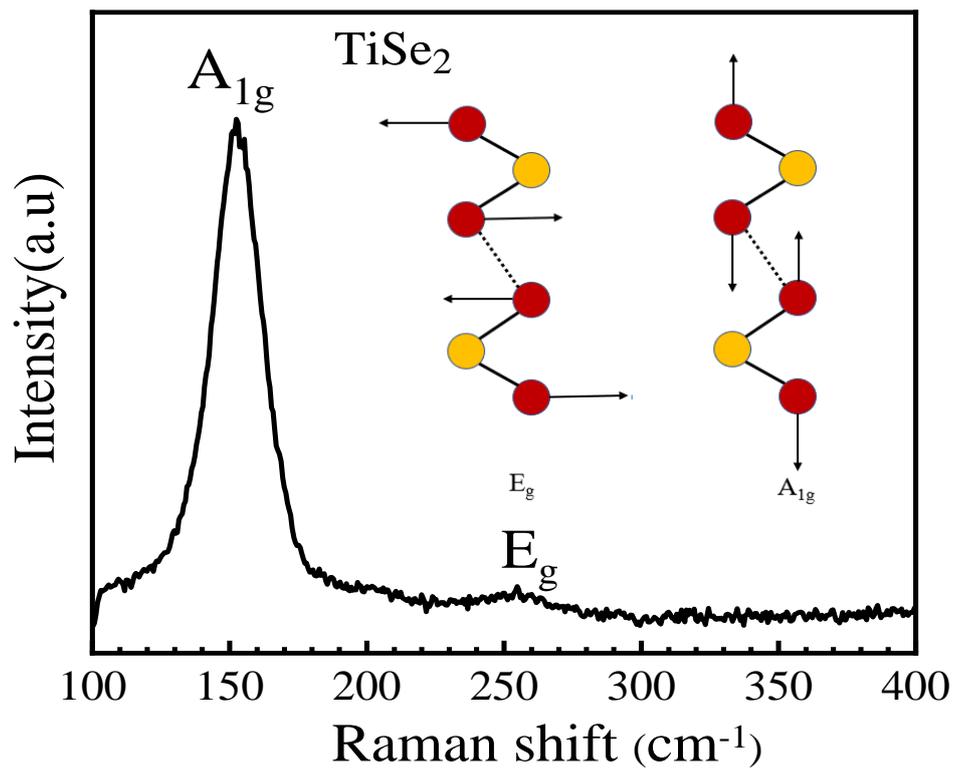

**Fig. 5(a)**

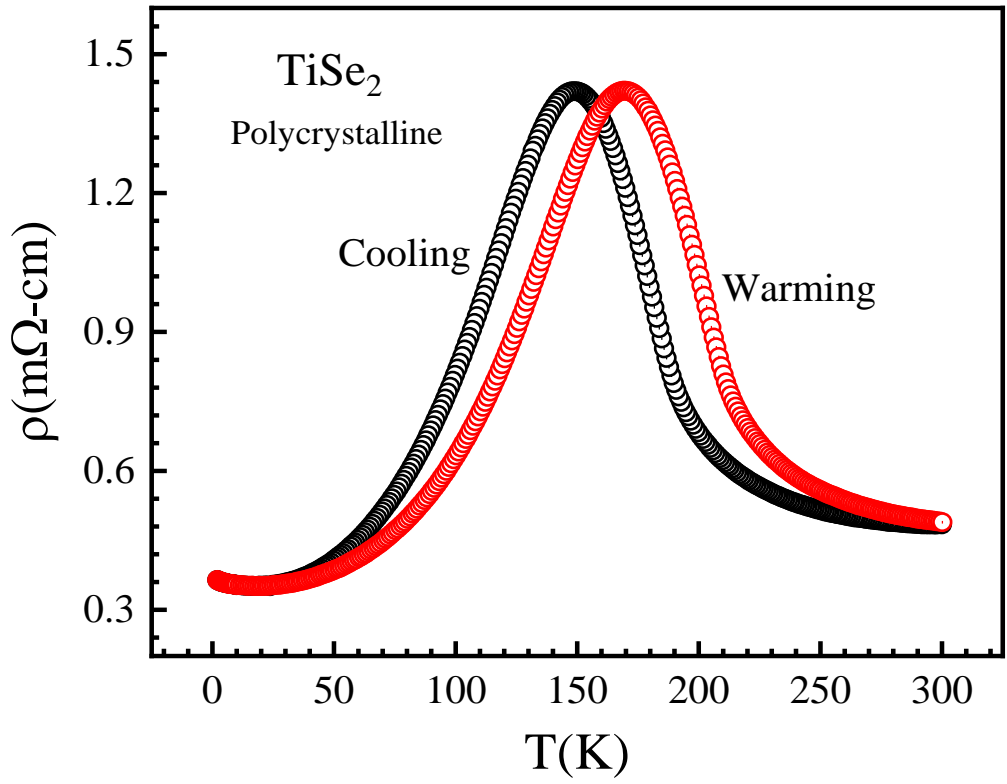

**Fig. 5(b)**

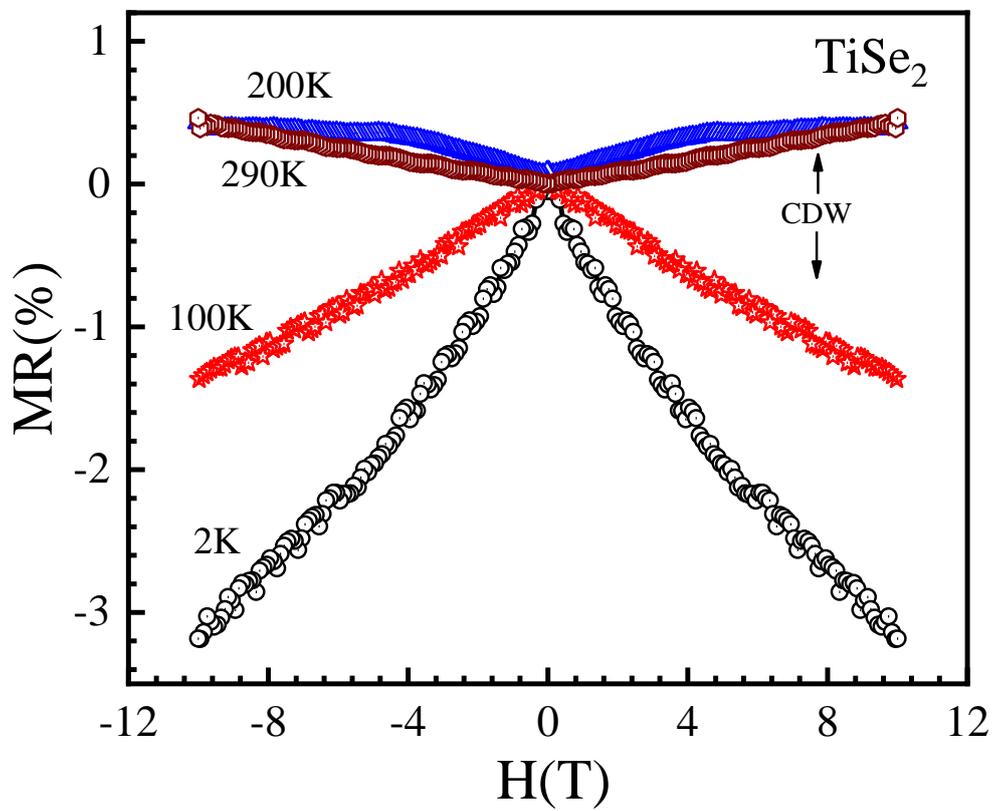